\documentclass[12pt]{article}
\usepackage{amsmath,amsfonts,graphicx,color,latexsym,cite}

\newcommand{\be}{\begin{equation}}
\newcommand{\ee}{\end{equation}}
\newcommand{\bea}{\begin{eqnarray}}
\newcommand{\eea}{\end{eqnarray}}

\newcommand{\half}{ {\frac 1 2}}
\newcommand{\bra}[1]{\langle #1 |}
\newcommand{\ket}[1]{| #1 \rangle}
\newcommand{\p}{\partial}
\newcommand{\refeq}[1]{(\ref{eq:#1})}
\newcommand{\ha}{\frac{1}{2}} \newcommand{\tha}{\tfrac{1}{2}}
\renewcommand{\epsilon}{\varepsilon}

\newcommand{\+}{\dagger}

\begin{document}


\begin{titlepage}
  \begin{flushright}
    \small hep-th/0302050 \\
    \small BROWN-HET-1341
\end{flushright}

\begin{center}
  \vspace{5mm}
{\Large \bf A note on $\alpha$-vacua and interacting field theory in de Sitter 
space}\\
\vspace{10mm}
Kevin Goldstein\footnote{\tt  kevin@het.brown.edu} and David A. Lowe\footnote{\tt  lowe@het.brown.edu}
\vspace{3mm}

{ \sl Department of Physics, \\ Brown University,\\Providence,  RI 02912, 
USA.}
\end{center}

\vskip 1cm

\begin{abstract}
Using an imaginary time formalism,  
we set up a consistent renormalizable perturbation theory of a scalar field in a
nontrivial $\alpha$ vacuum in de Sitter space. Although one
representation of the effective action involves non-local interactions
between anti-podal points, we argue the theory leads to causal
physics when continued to real-time, 
and we prove a spectral theorem for the interacting two-point
function.  We construct the renormalized stress energy tensor and show
this develops no imaginary part at leading order in the interactions,
consistent with stability. 
\end{abstract}

\end{titlepage}

\section{Introduction}

A common problem in formulating quantum field theory on a curved
background is ambiguity in the choice of vacuum. In de Sitter space there
is a one-parameter family of vacua invariant under the de Sitter
group, which have been dubbed the
$\alpha$ vacua \cite{Chernikov:1968zm,Tagirov:1973vv,Mottola:1985ar,Allen:1985ux}.
These vacua are perfectly self-consistent in the context of free
theories.  It has long been suggested that the only physically
sensible vacuum is the Euclidean (a.k.a. Bunch-Davies) vacuum. 

One reason for this choice is that the free propagators in the
Euclidean vacuum exhibit a Hadamard singularity, which matches with
what is expected in the flat space limit
\cite{Kay:1991mu,Wald:1995yp}. 
However the physical motivation for restriction to
Hadamard singular propagators is obscure in the context of interacting
quantum field theory, and certainly nothing
appears to go wrong with the $\alpha$-vacuum propagators at the free level. In
particular, as shown in \cite{Allen:1985ux} the commutator Green function is
vanishing at spacelike separations in an $\alpha$-vacuum and in fact
is independent of $\alpha$.

The Green functions in a nontrivial $\alpha$ vacuum exhibit singular
correlations between anti-podal points. Of course since the commutator
is compatible with locality this does not lead to acausal propagation
of information. However some
authors have suggested that once interactions are included the
$\alpha$-vacua do not lead to a sensible perturbative expansion of
Green functions. Banks et al. \cite{Banks:2002nv} have argued non-local counter-terms
render the effective action inconsistent. Einhorn et
al. \cite{Einhorn:2002nu} have argued
that $\alpha$ vacuum correlation functions are non-analytic and
conclude that they are physically unacceptable. 
Related arguments are made in
Kaloper et al. \cite{Kaloper:2002cs}, 
who argue that because an Unruh detector is not in thermal
equilibrium in an $\alpha$ vacuum, thermalization will lead to
decay to a Euclidean vacuum.

This issue has direct bearing on the theory of inflation. The
conventional view of inflation places the inflaton in the Euclidean
vacuum. However, as
emphasized in
\cite{Danielsson:2002kx,Danielsson:2002qh,Goldstein:2002fc} , 
the initial conditions for inflation
may place the inflaton in a non-trivial $\alpha$-vacuum, see also
\cite{Shankaranarayanan:2002ax} for earlier work in this direction. This has a
potentially large effect on the predictions for the CMB
spectrum, see for example
\cite{Danielsson:2002kx,Danielsson:2002qh,Goldstein:2002fc,Bergstrom:2002yd}
and references therein.
Furthermore, if there is a residual value of $\alpha$ today,
there are many interesting predictions for other observable quantities
such as cosmic rays, that we consider in more detail in \cite{lowe3}.

In this paper we show in an imaginary time formulation 
the $\alpha$-vacua do indeed have a well-defined
perturbative expansion, that yields finite renormalized amplitudes in
a conventional manner. This goes a long way to refuting some of the
objections raised in
\cite{Banks:2002nv,Einhorn:2002nu,Kaloper:2002cs}, see also
\cite{Danielsson:2002mb} for discussion of consistency of $\alpha$-vacua.

We begin in section 2 by reviewing the free field results of
\cite{Chernikov:1968zm,Tagirov:1973vv,Mottola:1985ar,Allen:1985ux}. 
In
particular, the $\alpha$-vacuum may be regarded as a squeezed state
created by a unitary operator $U$ acting on
the Euclidean vacuum. This idea will be central to the formalism
we develop. This leads to a generalized Wick's theorem, which allows
us to expand any free $\alpha$-vacuum Green function in terms of
products of Euclidean vacuum two-point functions. In section 3 we
describe the interacting theory in an imaginary time formalism. 
In particular, in interaction picture, we show the effective
Lagrangian becomes non-local when $U$ is commuted through the
fields. 
We show that UV divergences in amplitudes satisfy a non-trivial factorization
relation which relates the coefficients of local counter-terms to
non-local ones. Once local counter-terms are fixed, non-local terms
are completely determined, which implies the theory is renormalizable
in the conventional sense. In section 4 we outline how to continue the
imaginary time amplitudes to real time. We carry this through in
detail for the interacting two-point function, and prove a spectral
theorem in this case. One immediate consequence is that even in the
interacting theory, the
expectation value in an $\alpha$-vacuum of the commutator of two
fields vanishes at spacelike separations, as required for causality of
local observables. In section 5 we use the Green function to define a
renormalized stress energy tensor, and we conclude in section 6.

\section{Free fields}

\label{sec:free}

Let us begin by reviewing the construction of the $\alpha$-vacua
\cite{Allen:1985ux,Mottola:1985ar,Chernikov:1968zm,Tagirov:1973vv}. We
will set the Hubble radius $1/H=1$ unless otherwise stated.
In general a free scalar  field has the mode expansion
\begin{equation}
  \label{eq:free1}
  \phi(x) = \sum_n a_n \phi_n(x) + a_n^\+\phi^*_n(x)
\end{equation}
where $\{\phi_n\}$ satisfy the Klein-Gordon equation and
$[a_n,a^\+_m]=\delta_{nm}$.
 The $\phi_n$ are complete and orthonormal with respect to the Klein-Gordon product
\be
(\phi_1(x), \phi_2 (x) ) = -i \int_\Sigma ( \phi_1 \p_\mu \phi_2^* -
\phi_2^* \p_\mu \phi_1 ) d\Sigma^\mu
\ee
with $\Sigma$ a spacelike slice. In this paper we consider scalars
with mass $m$ and $R \phi^2$ coupling $\xi$. In a fixed de Sitter
background we can absorb $\xi$ into a redefinition of $m$, so we drop
$\xi$ from now on.

The vacuum state is characterized by
\begin{equation}
  \label{eq:free3}
  a_n\ket{\Omega_a}=0~.
\end{equation}
In general we can expand in another set of modes
\begin{equation}
  \label{eq:free81}
  \phi(x) = \sum_n b_n \phi_n^\alpha(x) + b_n^\+{\phi^\alpha_n}^*(x)
\end{equation}
related to the first by a Bogoliubov transformation. A new vacuum state
is defined by
\begin{equation}
  \label{eq:free7}
  b_n\ket{\Omega_b}=0~.
\end{equation}

As mentioned in dS space there is a one complex-parameter family of the dS
invariant vacua, dubbed the alpha vacua, $\ket{\alpha}$.  One of these, the
Euclidean or Bunch-Davies vacuum, $\ket{E}$, is defined by using mode
functions obtained by analytically
continuing mode functions regular on the lower half of the
Euclidean sphere. 
The Euclidean modes 
can be chosen such that $\phi^E_n(x)^*= \phi^E_n(\bar x)$, where $\bar x$ is 
the antipode of $x$, ($\bar x \equiv - x$). See, for example,
\cite{Mottola:1985ar,Bousso:2001mw} 
for explicit expressions of
these mode functions.

The modes of an arbitrary $\alpha$-vacuum general are related
to the Euclidean ones by a mode number independent Bogoliubov transformation,
\begin{equation}
  \label{eq:free4}
  \phi^\alpha_n = N_\alpha(\phi^E_n+e^\alpha{\phi_n^E}^*)
\end{equation}
where $N_\alpha\equiv(1-\exp (\alpha+\alpha^*))^{-\ha}$ and we require that 
${\rm Re}(\alpha)<0$. The Euclidean vacuum corresponds to $\alpha =
-\infty$.

In terms of creation and annihilation operators
\begin{equation}
  \label{eq:free5}
  b_n = N_\alpha(a_n-e^{\alpha*}{a_n^\+})
\end{equation}
where $b$ and $a$ are operators satisfying (\ref{eq:free3}) with
respect to the $\alpha$-vacuum and Euclidean 
respectively. This transformation can be implemented using a
unitary operator
\begin{equation}
  \label{eq:free6}
  b_n = {\cal U}a_n {\cal U}^\+
\end{equation}
where
\begin{equation}
  \label{eq:free8}
  {\cal U}_\alpha\equiv \exp(\sum_n c_\alpha {a^\+_n}^2-c^*_\alpha a_n^2)
\end{equation}
\begin{equation}
  \label{eq:free9}
    c_\alpha \equiv \frac{1}{4}\exp(-i{\rm Im}(\alpha))\log \tanh(-{\rm Re}(\alpha)/2)
\end{equation}
and we use the standard Taylor expansion of the exponential to define
the ordering.
The vacua are related by
\begin{equation}
  \label{eq:free10}
  \ket{\alpha}={\cal U}_\alpha\ket{E}
\end{equation}
since
\begin{equation}
  \label{eq:free11}
  b_n {\cal U} \ket{E} = {\cal U} a_n \ket{E} = 0~.
\end{equation}
From this perspective the $\alpha$-vacuum may be viewed as a {\it squeezed
state} on top of the usual Euclidean vacuum.

\subsection{Generalized Wick's theorem}

The Fock space built on the $\alpha$ vacuum is not unitary equivalent
to that of the Euclidean vacuum in general, because the unitary
transformation mixes positive and negative frequencies. 
However, as far as quantum
field theory in a fixed de Sitter background goes,
this unitary transformation leaves the complete 
set of physical observables
invariant. For our purposes, we take these observables to be finite
time Green functions, from which one may obtain $S$-matrix elements as
described in \cite{Dewitt:1975ys,Birrell:1982ix}.
All this unitary transformation does is to mix these observables up in a
non-local way, as we explain in more detail later in this section. 
This is the underlying reason that the interacting
$\alpha$-vacuum theory is consistent. 

If we were only considering the $\phi$ field on its own this would be
the end of the story. However if we wish to view $\phi$ as the
inflaton, physics dictates that $\phi$ should be
locally coupled to other fields. Thus we are interested in correlators
of the field $\phi$ with respect to the $\alpha$ vacuum. The
conjugated field ${\cal U}\phi(x){\cal
U}^\dagger$ on the
other hand,
would yield correlators in the $\alpha$-vacuum 
identical to the usual Euclidean vacuum correlators of $\phi$, but
would be coupled non-locally to other fields.

Since the unitary transformation
involves modes of arbitrarily high frequency (up to some physical cutoff) the
systematics of renormalizable perturbation theory will be quite
different from the usual Euclidean vacuum perturbation theory
\cite{Adler:1972qq,Adler:1973ty,Drummond:1975yc,Drummond:1979dg,Drummond:1979dn,Drummond:1979uy}. It will be our goal in the rest of this paper
to elaborate on renormalizable perturbation theory in the
$\alpha$-vacuum.

The correlators of interest take the form of expectation values of
products of fields $\phi$ with
respect to the state
$\ket{\alpha}$, or equivalently as conjugated fields
$\tilde\phi\equiv {\cal U}^\dagger\phi{\cal U}$ with respect to  $\ket{E}$
\begin{equation}
  \label{eq:free12}
  \begin{array}{lcl}
    \bra{\alpha}\phi(x_1) \phi(x_2) \ldots \phi(x_n)\ket{\alpha} &=&
    \bra{E}{\cal U}^\+
    \phi(x_1) {\cal U}{\cal U}^\+\phi(x_2){\cal U} {\cal U}^\+\ldots {\cal U}{\cal 
U}^\+\phi(x_n)
    {\cal U}\ket{E}\\
    &=&
    \bra{E}\tilde\phi(x_1) \tilde\phi(x_2) \ldots \tilde\phi(x_n)\ket{E}~.
  \end{array}
\end{equation}
Now, letting 
$\gamma=e^\alpha$ (so that $|\gamma|<1$),
\begin{equation}
  \label{eq:mode}
  \begin{array}{ll}
  \tilde\phi(x) & ={\cal U}^\dagger\phi(x){\cal U}\\
  &= {\cal U}^\+
  \left(
    {\displaystyle\sum_n} \phi^\alpha_n(x) b_n + \phi_n^{\alpha*}(x) 
b_n^\+\right){\cal U} \\
  &={ \displaystyle\sum_n} \phi^\alpha_n (x)a_n + \phi_n^{\alpha*}(x) a_n^\+ 
\\
    &= N_\alpha {\displaystyle\sum_n}
  \left(\phi^E_n(x)+\gamma{\phi_n^E}(\bar x)\right) a_n
   + \left({\phi^E_n}(x)+\gamma{\phi_n^E}(\bar x)\right)^* a_n^\+  \\
& = N_\alpha \left( \phi_0(x) + \phi_1(x) \right)
  \end{array}
\end{equation}
where we have defined
\be
\phi_0(x) \equiv \phi(x) ~, \qquad \phi_1(x) \equiv \sum_n \gamma
\phi_n^E(\bar x) a_n + \gamma^* {\phi_n^{E}}^*(\bar x) a_n^\dagger
\ee
If $\gamma$ is real, then $\tilde \phi(x)$ is simply a
linear combination of $\phi(x)$ and $\phi(\bar x)$. For $\gamma$
complex this isn't quite true, but the additional phases are simple to
keep track of.

Using these relations we can express any $\alpha$-vacuum correlator in 
terms of a sum of Euclidean vacuum correlators, giving us a
generalized Wick's theorem.
The simplest example is
\begin{equation}
  \label{eq:free13}
  \begin{array}{lll}
  \langle\alpha|\phi(x)\phi(y)|\alpha\rangle &=&  N_\alpha^2 \left(G_E(x,y)+|\gamma|^2 G_E(\bar x,\bar y)+\gamma G_E(\bar 
x,y)+\gamma^{*} G_E(x,\bar y) \right) \\
  &\equiv& G_\alpha(x,y)
  \end{array}
\end{equation}
where $G_E(x,y)\equiv \bra{E}\phi(x)\phi(y)\ket{E}$ is the Wightman
function on the Euclidean vacuum.  
It is convenient to introduce a two index notation,
\be
\label{twoind}
G_\alpha(x,y) = \sum_{i,j=0,1} G_{ij}(x,y)
\ee
where
\be
\label{indprop}
\begin{matrix}
G_{00} (x,y) &  = &  N_\alpha^2 G_E (x,y)~, \qquad &  G_{10} (x,y) & = &
N_\alpha^2 \gamma G_E (\bar x,y)~, \cr G_{01} (x,y) & = &
N_\alpha^2 \gamma^* G_E (x,\bar y)~, \qquad & G_{11}(x,y) & =&  N_\alpha^2
|\gamma|^2 G_E (\bar x,\bar y) \cr \end{matrix}
\ee
which we will use later.

The Wightman function
diverges when $x$ and $y$ are null separated.  As we can see from
\refeq{free13}, for the $\alpha$-vacua, there are additional divergences
when one point is null separated with the antipode of another.
This feature has led many to
consider the $\alpha$-vacua unphysical 
\cite{Einhorn:2002nu,Banks:2002nv,Kaloper:2002cs}.

\begin{figure}[htbp]
  \label{prop1}
    \begin{center}
         \includegraphics[width=0.75\textwidth]{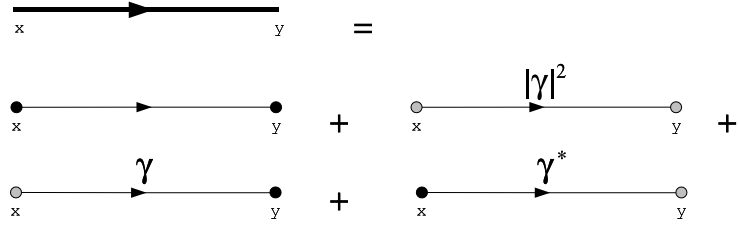}
    \end{center}
    \caption{\it Feynman diagram for \refeq{free13}.
      The thick line represents the $\alpha$-vacuum propagator $G_\alpha(x,y)$.  Thin
      lines represent Euclidean vacuum propagators with a factor of
      $N_\alpha^2$, and the other factors are shown explicitly. 
Grey dots denote points that appear in propagators as antipodes.  We have defined
      $\gamma=e^\alpha$.}
\end{figure}

Any correlation function of the form \refeq{free12}
in the free theory
can be found in terms of products of Green's functions by
normal-ordering the creation and annihilation operators, and retaining
the fully contracted terms. For example,
\begin{equation}
  \label{eq:free14}
  \begin{array}{lll}
   \bra{\alpha}\phi(x) \phi(y) \phi(z) \phi(w)\ket{\alpha}\\ = 
G_\alpha(x,y)G_\alpha(z,w)+G_\alpha(x,z)G_\alpha(y,w)+G_\alpha(x,w)G_\alpha(y,z)\\
     \end{array}
\end{equation}
where we have in mind using  (\ref{eq:free13}) to expand in terms of the Euclidean
vacuum Green's functions. Note it is important the ordering of the
arguments of the Green's functions is inherited from the ordering in
the operator expression on the left-hand side. This is because we are
stating the generalized Wick's theorem in the form of Wightman
functions rather than the usual form with time-ordered Green's
functions  \cite{Wick:1950ee}. The expansion of operator products in
free field theory using Wightman functions actually predates Wick's
theorem \cite{houriet}.
The theorem may be extended
to time-ordered expectation values by replacing the Wightman functions
with time-ordered two-point functions. 

To convert some diagram written in terms of the $G_\alpha$'s into one
in terms of Euclidean propagators, replace each thick line with a sum
of 4 thin
ones. 
To find the coefficient of each term just count up the number of grey dots,
noting their orientation with respect to the arrows. This is written
more compactly using the two index notation (\ref{twoind}), with an
index $i=0,1$ appearing at the end of each propagator, and all indices
summed over.

\section{Interacting fields}

So far all we have said is valid regardless of whether we work on de
Sitter space, or its Euclidean continuation, the four-sphere. Once we
introduce interactions, however, the choice of Lorentzian versus
Euclidean signature has a major impact on the formalism used to setup
the perturbative expansion. This is familiar from finite temperature
field theory where one has an imaginary time formalism \cite{abrikosov,Kapusta:1989tk}
or alternatively one can use a formulation in terms of real time
propagators at the price of doubling the number of fields
\cite{Takahasi:1975zn,Landsman:1987uw}. 

Describing interacting fields in curved spacetime with event horizons
using Lorentzian signature formalism is problematic. Inevitably one
must deal with propagators on opposite sides of the horizon, and
this leads to ambiguities in the formulation of Feynman rules. The
same problem exists for spacetimes with cosmological horizons, as
would arise if one attempted to quantize a field in de Sitter space in
the static coordinate patch. To avoid these issues we formulate
interacting field theory using the imaginary time, or Euclidean
continuation, as advocated in \cite{Hawking:1981ng}. 
Eventually we have in mind defining real-time ordered correlators
which may be used to construct in-out $S$-matrix elements as described 
in \cite{Dewitt:1975ys,Birrell:1982ix}. 

In the Euclidean vacuum, this problem has been much studied in the
literature
\cite{Adler:1972qq,Adler:1973ty,Drummond:1975yc,Drummond:1979dg,Drummond:1979dn,Drummond:1979uy,Harris:1994ge}
and corresponds to doing field theory on $S^4$.
Our strategy will be to 
use (\ref{eq:free10}) to define the interacting $\alpha$-vacuum, and
construct physical observables as correlators of the field $\phi$
which couple locally to physical sources. To 
evaluate these observables we set up the perturbation theory on the
Euclidean sphere as a non-local field theory in terms of
$\tilde \phi = {\cal U}^\dagger\phi(x){\cal U}$. We define these fields
as interaction picture fields, and will work with the standard methods
of canonical quantization.

The interacting part
of the non-local
action for $\tilde \phi$ is obtained by conjugating the
local bare interaction terms written in terms of the $\phi$ fields with the
operator $U$, or more precisely
\be
\label{nonloc}
T_{\tau} e^{i S_{non-local}^{int}(\tilde \phi) } \equiv {\cal U}^\dagger T_{\tau}
e^{i S_{local}^{int}( \phi) } {\cal U}
\ee
where $T_\tau$ denotes imaginary time ordering. The actions are obtained
by integrating the lagrangian density describing the interactions
(which we assume to be polynomial in $\phi$) over the Euclidean
sphere. 
We note that when
(\ref{eq:mode}) is substituted into this expression, the anti-podal
components $\phi_1(x)$ are to be ordered according to $x$ rather than
$\bar x$, since the ordering is to be inherited from the right-hand
side of (\ref{nonloc}).
The determination of any correlation function of $\phi$'s in an
$\alpha$-vacuum then
reduces to a standard Euclidean vacuum correlator computation, albeit
with some terms involving fields with unconventional time
ordering. That is, we expand (\ref{nonloc}) perturbatively in the
interactions, generating a sum of correlators of $\tilde \phi$ with
respect to the Euclidean vacuum. These may then be evaluated using the
generalized Wick's theorem of the previous section. 

Working on the Euclidean sphere has the advantage that a wide range of 
sensible cutoffs are available. For example one can choose dimensional 
regularization as in
\cite{Drummond:1975yc,Drummond:1979dg,Drummond:1979dn,Drummond:1979uy}, 
or simply a mode cutoff corresponding to a cutoff on angular momentum
on the 4-sphere. Pauli-Villars is another option, as is
point-splitting (with spherically symmetric averaging assumed to restore 
the symmetries), or zeta-function regularization \cite{Birrell:1982ix,Bunch:1978yq,Dowker:1976tf}. 
Little of what we say in the present work is
dependent on a particular choice of cut-off.

Let us comment further on the form of the correlators. 
The normalized Green functions in the $\alpha$-vacuum take the form
\bea
\label{green}
G(x_1, \cdots, x_n) & = &
 \frac {\langle {\cal U}^\dagger T_\tau  \phi(x_1) \cdots \phi(x_n)
  e^{i S^{int}_{local}(\phi)} {\cal U} \rangle }
{\langle {\cal U}^\dagger T_\tau   e^{i S^{int}_{local}(\phi)} {\cal U}
  \rangle } \nonumber \\
& = &
\frac {\langle T_\tau  \tilde \phi(x_{1}) \cdots \tilde \phi(x_{n}) e^{i
      S^{int}_{non-local}(\tilde \phi)} \rangle }
{\langle T_\tau   e^{i S^{int}_{non-local}(\tilde \phi)} \rangle } ~.
\eea
In imaginary time we cannot take an asymptotic limit
where interactions turn off, which is important in the usual
definition of the $S$-matrix to obtain the interacting vacuum. Instead 
we will simply compute correlators with respect to the free vacuum as
in (\ref{green}). As usual the denominator in (\ref{green}) implies we drop
disconnected diagrams when we compute Green functions. \footnote{By
  {\it disconnected} we mean diagrams not connected to
  external lines.} Because we are
not taking an LSZ type limit, the relevant Green functions correspond
to unamputated diagrams. We discuss continuation to real-time
amplitudes in the next section.

\begin{figure}[htbp]
  \label{prop2}
    \begin{center}
         \includegraphics[width=.75\textwidth]{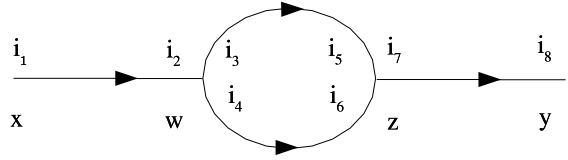}
    \end{center}
    \caption{\it Feynman diagram for propagator in $\lambda
      \phi^3$. The indices $i_k$ label end-points of the propagators
      $G_{i j}$. All indices are to be summed over.
      Vertex factors carry no $i_k$ dependence.}
\end{figure}

Let us now go through an example to illustrate the renormalization of
mass in $\lambda \phi^3$. The relevant Feynman diagram in position
space is shown in
(\ref{prop2}). The vertex $V_{ijk} =1 $ for all $i,j,k$, so the
amplitude is
\be
\label{massct}
A = \sum_{ \{ i_k =0,1\} }\int \sqrt{{\rm det} g_{\mu\nu}(w)} dw  \sqrt{{\rm det} g_{\mu\nu}(z)}
dz G^F_{i_1 i_2}(x,w) G^F_{i_3 i_4} (w,z)
G^F_{i_5,i_6}(w,z) G^F_{i_7 i_8} (z,y)
\ee
where $G^F$ is the time-ordered Green function
\be
G^F(x,y) = \theta(x^0-y^0) G_\alpha(x,y) + \theta(y^0-x^0)
G_\alpha(y,x)~,
\ee
and $x^0$ is the imaginary time coordinate on the sphere.
UV divergences arise when $w \to z$ or $w \to \bar z$.   In these limits
the propagator has the form
\be
\lim_{w\to z} G^F(w,z) = N_\alpha^2 (1+ |\gamma|^2)
\frac{1}{(w-z)^2}~,\qquad 
\lim_{w\to \bar z} G^F(w,z) = N_\alpha^2 (\gamma+\gamma^*)
\frac{1}{(w-\bar z)^2}
\ee
in locally Minkowski coordinates.  The UV divergent part of the
amplitude is then

\be
A_{UV} = \frac{\delta m^2 }{2}\int  \sqrt{{\rm det} g_{\mu\nu}(z)} dz \langle {\cal
  U}^\dagger \phi(x) \phi(z)^2 \phi(y) {\cal
  U} \rangle
\ee
where $\delta m^2$ is the cutoff dependent counter-term. If we adopt a
simple point-splitting regularization, this is given by
\bea 
\delta m^2 &=& \left( \int_{|w-z|\geq \epsilon}+\int_{|w-\bar z| \geq
    \epsilon }\right)
  \sqrt{{\rm det} g_{\mu\nu}(w)} dw \sum_{i_k} G_{i_3 i_4}(w,z) G_{i_5
    i_6}(w,z)  \nonumber \\
& \propto & N_\alpha^4 \left( (1+|\gamma|^2) ^2+ (\gamma+\gamma^*)^2 \right) \log
\epsilon ~.
\eea
We conclude
therefore that the counter-term is indeed simply a local mass
counter-term when expressed in terms of $\phi$ variables, but appears
non-local when written in terms of $\tilde \phi$ variables.

It is perhaps worthwhile to highlight the difference between our
computation and a similar computation of \cite{Banks:2002nv}.
Reference \cite{Banks:2002nv} assumed the basic vertex  was local. However in our formulation of the
$\alpha$-vacuum field theory, the vertex takes the form ${\cal
  U}^\dagger \phi(z)^3  {\cal
  U}$ which looks non-local when we expand this out in terms of the
fields $\phi_0(z)$ and $\phi_1(z)$, since we can view $\phi_1$ as localized
at $\bar z$. This non-locality is exactly what we need to make sense
of the non-local counter-terms encountered in
\cite{Banks:2002nv}. When all the diagrams are included the
coefficients of the non-local counter-terms are such that they arise
from the local counter-term $\half \delta m^2 \phi(x)^2$, prior to conjugation
by the ${\cal U}$'s. This implies the $\alpha$-vacuum perturbation
theory is rendered finite by the same number of renormalization
conditions as the corresponding Euclidean vacuum theory.

\section{Real-time correlators and causality}

We now discuss how to continue the imaginary time Green functions to
real time. In general this procedure is rather difficult as the
analytic continuation is not uniquely defined.  
One encounters similar problems in the formulation of Minkowski space
quantum field theory at finite temperature
\cite{Evans:1994cs,Evans:1992ky,Evans:1990hy,Guerin:1994ik}. 
There the analytic continuation from imaginary time to real-time, with
the extra condition that propagators be analytic in the lower half
frequency plane, computes retarded Green
functions. Retarded and advanced Green functions may be further
combined to give real time-ordered Green
functions. This procedure of determining propagators by analytic
continuation
 can be avoided by working
with the real-time thermo-field expansion of \cite{Takahasi:1975zn},
where the field content is doubled. In thermo-field theory, one also can
formulate a non-perturbative path integral definition of the theory 
using a non-trivial
real-time integration contour. It would be interesting to see if the
$\alpha$-vacuum theory could be formulated in an analogous way.
We will not develop that
here, but content ourselves for the moment with the perturbative
description of the theory described in the previous section. We will
use these results to obtain the analytic continuation to real-time 
of the general
interacting two-point function.

The general two-point function $ G$ in the interacting theory is
\be 
 G(x,y ) = \langle \Omega | \phi(x) \phi(y) | \Omega\rangle 
\ee
and is perturbatively defined by (\ref{green}). We use $\phi(x)$
to denote Heisenberg operators in this section. We can insert a complete set of states
to obtain
\be
G(x,y)= \sum_{\chi,n} \langle \Omega | \phi(x) | \chi, n \rangle
\langle \chi,n | \phi(y) | \Omega \rangle
\ee
where $| \chi,n\rangle$ denotes a scalar state with quantum numbers
$n$. We now use de Sitter symmetry to translate $\phi(x) = T \phi(0)
T^{-1}$, where $T$ is a de Sitter translation. $  | \Omega
\rangle $ is invariant under this translation. Usually one would
assume $ T^{-1}| \chi, n \rangle = \phi^E_{n}(x) |
\chi \rangle$,
but as we have learnt, invariance under the subgroup of the de Sitter
group continuously connected to the identity, in general only implies
$ T^{-1}| \chi, n \rangle = N_{\alpha(\chi) }( \phi_{n}^E(x) +
e^{\alpha(\chi)} \phi_{n}^{E*}( x)) | \chi \rangle$, where $\alpha$ is now
a function of the state $\chi$, and $\phi_n(x)$ is the generalization
of the modes to a scalar field of general mass
$m(\chi)$.
This implies
\be
G(x,y)= \sum_{\chi} \rho( \chi) G_{\alpha(\chi)}(m(\chi); x,y)
\ee
where $\rho$ is positive semi-definite. We can choose to parameterize this
instead as
\be
\label{spectral}
G(x,y)= \int_{m_{min}}^\infty dm \int d\alpha d\alpha^* \rho( m,\alpha) G_{\alpha}(m; x,y)
\ee
where $G_\alpha(m;x,y) $ 
is the generalization of (\ref{indprop}) to a free field of mass $m$.
In \cite{Bros:1994dn} it was argued $m_{min}=3/2$ for the theory in
the Euclidean vacuum. The $m>3/2$ scalar representations of the de
Sitter group are known as the principal series, and only these have a
smooth limit to representations of the Poincare group as $H \to 0$
\cite{Gazeau:1999mi}. The $0<m<3/2$ representations are known as the
complementary series.
There do not appear to be any
obvious problems with quantizing fields with $0<m<3/2$. For example,
the conformally 
coupled free scalar ($m=\sqrt{2}$), is related by a conformal
transformation to a massless field in flat space. 
Since we are often interested in
fields with $m<3/2$, we include the complementary series in our space
of allowed states, so take $m_{min}=0$.

The $i \epsilon$ prescriptions for the propagators $G_{jk}$ are
defined in appendix B, which allows the $ G$ to be continued
to a function regular on the Lorentzian section. This prescription is fixed
by imposing the boundary condition that each component of the two-point function
$G_{\alpha}(m; x,y)$ match the free Wightman propagators constructed by Mottola
and Allen \cite{Mottola:1985ar,Allen:1985ux}.
Appropriate
linear combinations of $ G(x,y)$  define the real-time retarded, advanced,
and time-order propagators. We note the complete propagator $ G$
is not analytic in the lower half $t$ plane (see appendix B for
notation), but it is built out of terms, each of which separately
enjoys analyticity in the upper or lower half $t$ plane.

Demonstrating causality of the interacting two-point function is now
trivial. We simply apply (\ref{spectral}) to the commutator of two
fields
\bea
\langle  [ \phi(x), \phi(y) ] \rangle_\alpha & =&   G(x,y) -
 G(y,x) \\
&=&  \int_0^\infty dm d\alpha d\alpha^* ~\rho(m,\alpha) ( G_{\alpha} (m; x,y) -
G_{\alpha} (m; y,x))  \nonumber \\
& = &  \int_0^\infty dm  d\alpha d\alpha^* ~\rho(m,\alpha ) ( G_E (m; x,y) -
G_E (m; y,x))  \nonumber 
\eea
which vanishes at spacelike separations of $x$ and $y$. Here we have
used the result of \cite{Allen:1985ux} that the commutator in the free
theory is independent of $\alpha$.

To sum up, we have defined a continuation of the general interacting
two-point function from imaginary time to real time, using a spectral
theorem and we have shown the real-time commutator function is causal
despite the apparent non-analyticity of the perturbative expansion. 

\section{Stress-Energy Tensor}
 
Numerous
techniques for calculating $\langle T_{\mu\nu}\rangle$ in general, and
in the Euclidean vacuum of de Sitter space in particular, are reviewed
in \cite{Birrell:1982ix,Bunch:1978yq,Dowker:1976tf} and references therein. 
Since  \cite{Bernard:1986vc}  considered the stress-energy tensor in
the $\alpha$-vacua some time ago, we mainly
quote their results.
The stress-energy tensor for a scalar field is given by 
\begin{equation}
  \label{eq:set1}
  T_{\mu\nu}= \phi_{,\mu}\phi_{,\nu}-
  \tha g_{\mu\nu}\phi^{,\alpha}\phi_{,\alpha}
  + g_{\mu\nu}V(\phi)
\end{equation}
where, for simplicity, we have set the $R\phi^2$ coupling to $0$.

In general, we can find the renormalized expectation value of
$T_{\mu\nu}$ for a non-interacting scalar field from the symmetric
Greens function $G^{(1)}_{xy}=\langle\{\phi_x,\phi_y \}\rangle$, as follows:
\begin{equation}
  \begin{array}{l}
 \langle T_{\mu\nu}(x)\rangle_{ren}^{free} = \\
  {\displaystyle \lim_{x',x''\rightarrow x}} 
  \left(\nabla_{\mu'}\nabla_{\nu''} 
  -\tha g_{\mu\nu}\nabla^{\gamma'}\nabla_{\gamma''}+ \tha m^2 g_{\mu\nu}\right)
  \tha \left( G^{(1)}(x',x'')- G_{ref}(x',x'')\right)
  \end{array}
\end{equation}
where $G_{ref}$ is a reference two-point function which removes the
singularities in $G^{(1)}$. 
Note the limit, and $G_{ref}$ must be chosen  to preserve 
covariance and ${T^{\mu\nu}}_{;\nu}=0$.

Consistent with  previous work cited above  \cite{Bernard:1986vc}  found
for a non-interacting  Euclidean vacuum \cite{Bunch:1978yq}, 
\begin{equation}
\label{eq:set2} 
\begin{array}{l}
\langle T_{\mu\nu} \rangle_{ren}^{free} = \\
  - \frac{ g_{\mu\nu}}{64\pi^2}
   \Big( m^2(m^2-2 H^2)
    \left(\psi\left(\tfrac{3}{2}-i\nu\right)+\psi\left(\tfrac{3}{2}+i\nu\right)
    +\ln\left(\frac{H^2}{\mu^2}\right)
     \right)   \\ 
   + \tfrac{8}{3}m^2 H^2-\tfrac{359}{180}H^4 \Big)  
\end{array}
\end{equation}
where  $H$ is Hubble's constant, $\nu=\sqrt{m^2/H^2-9/4}$ and
$\mu$ is some mass renormalization scale.
$\langle T_{\mu\nu}\rangle$ for a general $\alpha$-vacuum, with a
non-interacting scalar field, has been found by by \cite{Bernard:1986vc} to be
\begin{equation}
\label{eq:set3} 
\begin{array}{l}
\langle T_{\mu\nu} \rangle_{ren}^{free} = \\
  - \frac{ g_{\mu\nu}}{64\pi^2}
  \frac{1+|\gamma|^2}{1-|\gamma|^2}\times \\
   \Big(
        m^2(m^2-2 H^2)
    \left(\psi\left(\tfrac{3}{2}-i\nu\right)+\psi\left(\tfrac{3}{2}+i\nu\right)
    +\ln\left(\frac{H^2}{\mu^2}\right)+\frac{\pi}{\cosh\pi\nu}\frac{\gamma+\gamma^*}{1+|\gamma|^2}
     \right) \\ 
   + \tfrac{8}{3}m^2 H^2-\tfrac{359}{180}H^4 \Big) ~.
\end{array}
\end{equation}
 The main difference between \refeq{set2} and \refeq{set3} is an
extra factor of $\frac{1+|\gamma|^2}{1-|\gamma|^2}$. The origin of
this constant can be seen from the short distance limit of
\refeq{free13} which gives $G_\alpha(x,x) \sim
\frac{1+|\gamma|^2}{1-|\gamma|^2} G_E(x,x)$. The $\alpha$-dependence
of the short distance singularity means  that our counter terms must be
$\alpha$-dependent. The fact that $\alpha$-dependent counter-terms are required for a
finite $\langle T_{\mu\nu}\rangle$ was viewed as problematic in
\cite{Kaloper:2002cs}. 
As we have already emphasized previously these are precisely
the sort of counter-terms we naturally expect. We emphasize both
\refeq{set2} and \refeq{set3} are proportional to $g_{\mu\nu}$ which
is covariantly constant, implying conservation of energy.

An important conclusion we draw from \refeq{set3} is that no imaginary part
appears in $T_{\mu \nu}$ (and hence the action at one-loop
order). This indicates the $\alpha$-vacuum is stable at this order. We
discuss the possibility of higher order instabilities in the conclusions.

Now to calculate the $ \langle T_{\mu\nu}(x)\rangle$ for the
interacting case we need to replace the free Green
function with the interacting one (and add in $\langle V_I (\phi)
\rangle$). 
The spectral representation (\ref{spectral}) then yields a
straightforward generalization of \refeq{set3}.

\section{Conclusions}

We have constructed a renormalizable perturbation theory for scalar
field amplitudes in an $\alpha$ vacuum using an imaginary time formulation. We have also shown the theory
is causal when continued to real-time, at the level of the two-point
function.  It remains an interesting open problem to use this
formalism to construct the higher order real-time correlators. Our
hope is this may be achieved by taking appropriate linear combinations
of the imaginary time amplitudes, with external legs continued to
real-time, as is the case for finite temperature field theory.

These results are of importance for the theory of
inflation, because if the inflationary phase sat in a general $\alpha$
vacuum, the amplitude and spectrum of cosmic microwave background
perturbations can be dramatically effected \cite{Goldstein:2002fc}. 
If the present day universe is likewise asymptoting toward a universe
dominated by positive cosmological constant, the asymptotic value of
$\alpha$ can produce observable effects today, and may be responsible
for a component of the diffuse cosmic ray flux (see
\cite{Starobinsky:2002rp} for a study of this effect).
We plan to develop further the phenomenology of these
vacua in future work. 

The formalism we have developed may also have
useful generalizations to computations in flat space in squeezed state
backgrounds. See \cite{Svozil:1994tq} for QED calculations in squeezed
state backgrounds.

Let us emphasize that our 
motivation for this study was to try to discover a
problem with the $\alpha$-vacuum, which would lead one to conclude the
Euclidean vacuum was unique. Thus far we have not found such a
problem, which raises the question whether we must think of $\alpha$
as a new cosmological constant fixed by initial conditions, or whether
dynamics leads to decay to the Euclidean vacuum at late times. 

One
hint that it might be the latter comes from the fact that
requiring an Unruh detector see a thermal distribution of particles
uniquely selects the Euclidean vacuum.  Thus demanding local
equilibrium (assuming no extra chemical potentials are turned
on) selects the Euclidean vacuum. However that raises the question
whether non-trivial $\alpha$ simply corresponds to a new chemical
potential needed to uniquely specify the scalar field
theory.\footnote{\cite{Bousso:2001mw} suggest $\alpha$ can be
  interpreted as a marginal deformation of the CFT in the context of
  dS/CFT \cite{Strominger:2001pn}.} 
This then introduces a new tunable parameter into the effective field
theory description
of inflation. In section 5 we found no imaginary part in the action at
the one-loop level, indicating that at leading order the
$\alpha$-vacua are stable, consistent with this latter interpretation.

In \cite{Goldstein:2002fc} we argued in more general cosmological
backgrounds, $\alpha$ should be tied to the
cosmological constant $\Lambda \sim H_{eff}^2 M_{Planck}^2$, by
$e^{\alpha(t)} \sim H_{eff}/M_{cutoff}$. For concreteness, let us take
$M_{cutoff} \sim 10^{16}$ GeV, around the GUT scale.
We find it intriguing 
that bounds on $\alpha_{today}$ from diffuse cosmic ray observations
\cite{Starobinsky:2002rp} ($e^\alpha \leq 10^{-6}
\frac{H_{today}}{M_{cutoff}} \frac{M_{Planck}}{M_{cutoff}}$ in our notation) place
it in within a few of orders of magnitude of 
the scale $H_{today}/M_{cutoff}$.\footnote{In
\cite{Starobinsky:2002rp} it was
  assumed photons (and possibly other Standard model fields) were in
  an analog of an $\alpha$ vacuum. It is perhaps more natural to
  assume only the inflaton(s) are in an $\alpha$ vacuum, which could substantially
  weaken the bounds of \cite{Starobinsky:2002rp}.} Furthermore
cosmic rays have currently been observed with energy up to $10^{11}$ GeV
\cite{Bird:1993yi}, with no obvious upper cutoff in sight, 
consistent with $\alpha$ vacuum predictions.

\bigskip
\centerline{\large \bf Acknowledgements}
\noindent
We thank R. Brandenberger, A. Jevicki, S. Theisen and the Harvard High
Energy theory group for helpful discussions.
This research is supported in part by DOE
grant DE-FE0291ER40688-Task A.

\appendix
\section{Appendix: Squeezed states}
\renewcommand{\theequation}{A.\arabic{equation}}
\setcounter{equation}{0}
We record some formulas useful in the manipulation of squeezed states.

\begin{equation}
  {\cal U}(\zeta)\equiv \exp(\tha (\bar\zeta a^2-\zeta {a^\dagger}^2))=e^A
\end{equation}

\begin{equation}
  e^A a e^{-A} = e^{{\cal L}_A} a
\end{equation}
where ${\cal L}_A B \equiv [A,B]$. Let $\zeta=\rho e^{i\phi}$, then

\begin{equation}
  \begin{array}{lcl}
[A,a] &=& \zeta a^\dagger = \rho e^{i\phi} a^\dagger \\

[A,a^\dagger] &=& \zeta^* a~.
  \end{array}
\end{equation}

This implies

\begin{equation}
  \begin{array}{lcl}
{\cal L}_A^{2n+1} a &=&  e^{i\phi} \rho^{2n+1}a^\dagger \\
{\cal L}_A^{2n} a &=&  \rho^{2n}a \\
  \end{array}
\end{equation}
so we obtain
\begin{equation}
  \begin{array}{lll}
    e^{A}a e^{-A}
    &=& a \sum\rho^{2n}/2n! + a^\dagger e^{i\phi}\sum \rho^{2n+1}/(2n+1)!\\
    &=&  a_k \cosh (\rho) +  a_k^\dagger e^{i\phi}\sinh (\rho)
  \end{array}
\end{equation}
and
\begin{equation}
  \label{eq:bog1}
  b_k = {\cal U}a_k {\cal U}^\dagger=N_\alpha(a_k - e^{\alpha^*} 
a_k^\dagger)
  = a_k \cosh (\rho) +  a_k^\dagger e^{i\phi}\sinh (\rho)~.
\end{equation}

Some other expressions that we use:
\begin{equation}
 e^{\alpha-\alpha^*}=e^{-2i\phi}  \Rightarrow \phi=-{\rm Im}(\alpha)
 \end{equation}
\begin{equation}
  e^{\alpha+\alpha^*}= \tanh^2 \rho \\
\end{equation}
\begin{equation}
\begin{array}{lll}
\rho &=& \tanh^{-1}e^{{\rm Re}(\alpha)}\\
&=& \ha \ln (\frac{1+e^{{\rm Re}(\alpha)}}{1-e^{{\rm Re}(\alpha)}})\\
&=& \ha \ln \tanh (-\tha{\rm Re}(\alpha))
\end{array}
\end{equation}
\begin{equation}
  \zeta = \tha e^{-i{\rm  Im}(\alpha)}\ln \tanh (-\tha{\rm Re}(\alpha))~.
\end{equation}

\section{Appendix: Some useful facts}
\renewcommand{\theequation}{B.\arabic{equation}}
\setcounter{equation}{0}

Global coordinates
\be
ds^2 = -dt^2 + \cosh^2 t d \Omega^2
\ee
where $d\Omega^2$ is the metric on the unit 3-sphere. We will work in
units where the Hubble radius is 1. Define Euclidean
vacuum using mode functions
\be
\psi_{klm} (x) = y_k(t) Y_{klm} (\Omega)
\ee
where $k,l,m$ label the complete set of scalar spherical harmonics on
$S^3$,
$-|l| \leq m \leq |l|$. The $y_k(t)$ may be expressed in terms of the
hypergeometric function ${}_2F_1$ \cite{Mottola:1985ar}. 
These are regular on the Euclidean
section, and may be analytically continued to functions regular on the
lower half $\zeta$ plane, $\zeta = i \sinh t$.  They have a branch
cut from $\zeta =1 $ to $\zeta= \infty$. 

Define linear combination
\be
\phi_{klm} = \frac{e^{i \pi k/2} } {\sqrt{2}} \left( e^{i \pi/4}
  \psi_{klm}(x) + e^{-i \pi /4} \psi_{kl-m} (x) \right)
\ee
This set of modes is the basis of the complete set of modes we will
use. They are orthonormal and positive norm, and satisfy
\bea
\phi_{klm} (\bar x) &=& \phi^*_{klm}(x) \nonumber \\
(\phi_{klm}, \phi_{k' l' m'} ) &= & \delta_{k k'} \delta_{l l'}
\delta_{m m'} \\
\eea

By definition the Euclidean vacuum Green function satisfies
\be
\label{eucg}
G_E(x,y) = \sum_n \phi_n(x) \phi^*_n(y)
\ee
where we have compressed the $k,l,m$ indices into the single index
$n$. 

It is useful to define $z(x,y) = X\cdot Y$ where $X$ and $Y$ are the
coordinates of points on 5d Minkowski space, where de Sitter can be
embedded as $-X_0^2+X_1^2+X_2^2+X_3^2+X_4^2=1$. For spacelike
separations $z<1$, for null separations $z=1$, and for timelike
separations $z>1$. We have in mind continuing $z$ to complex values
for which the relation to geodesic distance breaks down.
Note also that $z(\bar x, y) = - z(x,y)$.
In terms of $z$, (\ref{eucg}) can be
written explicitly as
\be
\label{hypergeo}
G_E(x,y) =\frac{\Gamma(3/2 + i \nu ) \Gamma(3/2-i \nu) }{ (4 \pi )^2  }
{}_2F_1\left(3/2 + i \nu, 3/2-i \nu,2; (1+z)/2\right)
\ee
This function has a pole at $z=1$ and a branch cut extending from $z=1$ 
along the positive real axis. The function is analytic in the
lower-half $z$ plane. When $z$ is real, $G_E(z)$ is real for $z<1$,
and develops an imaginary part for $z>1$. The sign of this imaginary
part changes as one moves across the branch cut.

To make (\ref{eucg}) well-defined for time-like separations, we must
specify an $i \epsilon$ prescription.  Near the singularity $z=1$, we
specify this in locally Minkowski coordinates $(t,\vec x)$ by \cite{Bousso:2001mw}
\be
G_E(x,x') \sim \frac{1}{(t-t'-i\epsilon)^2 -|\vec x-\vec x'|^2 }~.
\ee

In the text we introduce the Green functions $G_{ij}(x,y)$. These are
likewise defined using $G_E$ but the $i \epsilon$ prescriptions are as
follows (for simplicity we set $\vec x$ and $\vec x'$ to 0):  
\bea
G_{00} (x,x')& = & N_\alpha^2 G_E(t-i\epsilon,t') \\
G_{10} (x,x') &=& N_\alpha^2 \gamma G_E (-t-i\epsilon,t') \\
 G_{01} (x,x') & = &
N_\alpha^2 \gamma^* G_E (-t+i \epsilon,t') \\
 G_{11}(x,x') & =&  N_\alpha^2
|\gamma|^2 G_E (t+i\epsilon,t') ~.
\eea

\bibliography{desitterqft.bib}

\providecommand{\href}[2]{#2}\begingroup\raggedright\begin{thebibliography}{10}

\bibitem{Chernikov:1968zm}
N.~A. Chernikov and E.~A. Tagirov, ``Quantum theory of scalar fields in de
  sitter space-time,'' {\em Annales Poincare Phys. Theor.} {\bf A9} (1968)
109.

\bibitem{Tagirov:1973vv}
E.~A. Tagirov, ``Consequences of field quantization in de sitter type
  cosmological models,'' {\em Ann. Phys.} {\bf 76} (1973)
561--579.

\bibitem{Mottola:1985ar}
E.~Mottola, ``Particle creation in de sitter space,'' {\em Phys. Rev.} {\bf
  D31} (1985)
754.

\bibitem{Allen:1985ux}
B.~Allen, ``Vacuum states in de sitter space,'' {\em Phys. Rev.} {\bf D32}
  (1985)
3136.

\bibitem{Kay:1991mu}
B.~S. Kay and R.~M. Wald, ``Theorems on the uniqueness and thermal properties
  of stationary, nonsingular, quasifree states on space-times with a bifurcate
  killing horizon,'' {\em Phys. Rept.} {\bf 207} (1991)
49--136.

\bibitem{Wald:1995yp}
R.~M. Wald, ``Quantum field theory in curved space-time and black hole
  thermodynamics,''. Chicago, USA: Univ. Pr. (1994) 205 p.

\bibitem{Banks:2002nv}
T.~Banks and L.~Mannelli, ``De sitter vacua, renormalization and locality,''
\href{http://www.arXiv.org/abs/hep-th/0209113}{{\tt hep-th/0209113}}.

\bibitem{Einhorn:2002nu}
M.~B. Einhorn and F.~Larsen, ``Interacting quantum field theory in de sitter
  vacua,''
\href{http://www.arXiv.org/abs/hep-th/0209159}{{\tt hep-th/0209159}}.

\bibitem{Kaloper:2002cs}
N.~Kaloper, M.~Kleban, A.~Lawrence, S.~Shenker, and L.~Susskind, ``Initial
  conditions for inflation,'' {\em JHEP} {\bf 11} (2002) 037,
\href{http://www.arXiv.org/abs/hep-th/0209231}{{\tt hep-th/0209231}}.

\bibitem{Danielsson:2002kx}
U.~H. Danielsson, ``A note on inflation and transplanckian physics,'' {\em
  Phys. Rev.} {\bf D66} (2002) 023511,
\href{http://www.arXiv.org/abs/hep-th/0203198}{{\tt hep-th/0203198}}.

\bibitem{Danielsson:2002qh}
U.~H. Danielsson, ``Inflation, holography and the choice of vacuum in de sitter
  space,'' {\em JHEP} {\bf 07} (2002) 040,
\href{http://www.arXiv.org/abs/hep-th/0205227}{{\tt hep-th/0205227}}.

\bibitem{Goldstein:2002fc}
K.~Goldstein and D.~A. Lowe, ``Initial state effects on the cosmic microwave
  background and trans-planckian physics,''
\href{http://www.arXiv.org/abs/hep-th/0208167}{{\tt hep-th/0208167}}.

\bibitem{Shankaranarayanan:2002ax}
S.~Shankaranarayanan, ``Is there an imprint of planck scale physics on
  inflationary cosmology?,'' {\em Class. Quant. Grav.} {\bf 20} (2003) 75--84,
\href{http://www.arXiv.org/abs/gr-qc/0203060}{{\tt gr-qc/0203060}}.

\bibitem{Bergstrom:2002yd}
L.~Bergstrom and U.~H. Danielsson, ``Can map and planck map planck physics?,''
  {\em JHEP} {\bf 12} (2002) 038,
\href{http://www.arXiv.org/abs/hep-th/0211006}{{\tt hep-th/0211006}}.

\bibitem{lowe3}
K.~Goldstein and D.~A. Lowe, ``to appear,''.

\bibitem{Danielsson:2002mb}
U.~H. Danielsson, ``On the consistency of de sitter vacua,'' {\em JHEP} {\bf
  12} (2002) 025,
\href{http://www.arXiv.org/abs/hep-th/0210058}{{\tt hep-th/0210058}}.

\bibitem{Bousso:2001mw}
R.~Bousso, A.~Maloney, and A.~Strominger, ``Conformal vacua and entropy in de
  sitter space,'' {\em Phys. Rev.} {\bf D65} (2002) 104039,
\href{http://www.arXiv.org/abs/hep-th/0112218}{{\tt hep-th/0112218}}.

\bibitem{Dewitt:1975ys}
B.~S. Dewitt, ``Quantum field theory in curved space-time,'' {\em Phys. Rept.}
  {\bf 19} (1975)
295--357.

\bibitem{Birrell:1982ix}
N.~D. Birrell and P.~C.~W. Davies, ``Quantum fields in curved space,''.
  Cambridge, Uk: Univ. Pr. ( 1982) 340p.

\bibitem{Adler:1972qq}
S.~L. Adler, ``Massless, euclidean quantum electrodynamics on the five-
  dimensional unit hypersphere,'' {\em Phys. Rev.} {\bf D6} (1972)
3445--3461.

\bibitem{Adler:1973ty}
S.~L. Adler, ``Massless electrodynamics on the five-dimensional unit
  hypersphere: An amplitude - integral formulation,'' {\em Phys. Rev.} {\bf D8}
  (1973)
2400--2418.

\bibitem{Drummond:1975yc}
I.~T. Drummond, ``Dimensional regularization of massless theories in spherical
  space-time,'' {\em Nucl. Phys.} {\bf B94} (1975)
115.

\bibitem{Drummond:1979dg}
I.~T. Drummond and G.~M. Shore, ``Conformal anomalies for interacting scalar
  fields in curved space-time,'' {\em Phys. Rev.} {\bf D19} (1979)
1134.

\bibitem{Drummond:1979dn}
I.~T. Drummond, ``Conformally invariant amplitudes and field theory in a
  space-time of constant curvature,'' {\em Phys. Rev.} {\bf D19} (1979)
1123.

\bibitem{Drummond:1979uy}
I.~T. Drummond and G.~M. Shore, ``Dimensional regularization of massless
  quantum electrodynamics in spherical space-time. 1,'' {\em Ann. Phys.} {\bf
  117} (1979)
89.

\bibitem{Wick:1950ee}
G.~C. Wick, ``The evaluation of the collision matrix,'' {\em Phys. Rev.} {\bf
  80} (1950)
268--272.

\bibitem{houriet}
A.~Houriet and A.~Kind, ``Classification invariante des termes de la matrice
  s,'' {\em Helv. Phys. Acta} {\bf 22} (1949) 319.

\bibitem{abrikosov}
A.~A. Abrikosov, L.~P. Gorkov, and I.~E. Dzyaloshinski, ``Methods of quantum
  field theory in statistical physics,''. Dover Publications (1975).

\bibitem{Kapusta:1989tk}
J.~I. Kapusta, ``Finite temperature field theory,''. Cambridge, Uk: Univ. Pr.
  (1989).

\bibitem{Takahasi:1975zn}
Y.~Takahasi and H.~Umezawa, ``Thermo field dynamics,'' {\em Collect. Phenom.}
  {\bf 2} (1975)
55--80.

\bibitem{Landsman:1987uw}
N.~P. Landsman and C.~G. van Weert, ``Real and imaginary time field theory at
  finite temperature and density,'' {\em Phys. Rept.} {\bf 145} (1987)
141.

\bibitem{Hawking:1981ng}
S.~W. Hawking, ``Interacting quantum fields around a black hole,'' {\em Commun.
  Math. Phys.} {\bf 80} (1981)
421.

\bibitem{Harris:1994ge}
B.~A. Harris and G.~C. Joshi, ``A new formulation of quantum field theory on
  s(4),'' {\em Int. J. Mod. Phys.} {\bf A9} (1994)
3245--3282.

\bibitem{Bunch:1978yq}
T.~S. Bunch and P.~C.~W. Davies, ``Quantum field theory in de sitter space:
  Renormalization by point splitting,'' {\em Proc. Roy. Soc. Lond.} {\bf A360}
  (1978)
117--134.

\bibitem{Dowker:1976tf}
J.~S. Dowker and R.~Critchley, ``Effective lagrangian and energy momentum
  tensor in de sitter space,'' {\em Phys. Rev.} {\bf D13} (1976)
3224.

\bibitem{Evans:1994cs}
T.~S. Evans, ``What is being calculated with thermal field theory?,''
\href{http://www.arXiv.org/abs/hep-ph/9404262}{{\tt hep-ph/9404262}}.

\bibitem{Evans:1992ky}
T.~S. Evans, ``N point finite temperature expectation values at real times,''
  {\em Nucl. Phys.} {\bf B374} (1992)
340--372.

\bibitem{Evans:1990hy}
T.~S. Evans, ``Spectral representation of three point functions at finite
  temperature,'' {\em Phys. Lett.} {\bf B252} (1990)
108--112.

\bibitem{Guerin:1994ik}
F.~Guerin, ``Retarded - advanced n point green functions in thermal field
  theories,'' {\em Nucl. Phys.} {\bf B432} (1994) 281--314,
\href{http://www.arXiv.org/abs/hep-ph/9306210}{{\tt hep-ph/9306210}}.

\bibitem{Bros:1994dn}
J.~Bros, U.~Moschella, and J.~P. Gazeau, ``Quantum field theory in the de
  sitter universe,'' {\em Phys. Rev. Lett.} {\bf 73} (1994)
1746--1749.

\bibitem{Gazeau:1999mi}
J.~P. Gazeau, J.~Renaud, and M.~V. Takook, ``Gupta-bleuler quantization for
  minimally coupled scalar fields in de sitter space,'' {\em Class. Quant.
  Grav.} {\bf 17} (2000) 1415--1434,
\href{http://www.arXiv.org/abs/gr-qc/9904023}{{\tt gr-qc/9904023}}.

\bibitem{Bernard:1986vc}
D.~Bernard and A.~Folacci, ``Hadamard function, stress tensor and de sitter
  space,'' {\em Phys. Rev.} {\bf D34} (1986)
2286.

\bibitem{Starobinsky:2002rp}
A.~A. Starobinsky and I.~I. Tkachev, ``Trans-planckian particle creation in
  cosmology and ultra- high energy cosmic rays,'' {\em JETP Lett.} {\bf 76}
  (2002) 235--239,
\href{http://www.arXiv.org/abs/astro-ph/0207572}{{\tt astro-ph/0207572}}.

\bibitem{Svozil:1994tq}
K.~Svozil, ``Quantum electrodynamics in the squeezed vacuum state: Feynman
  rules and corrections to the electron mass and anomalous magnetic moment,''
\href{http://www.arXiv.org/abs/hep-ph/9402316}{{\tt hep-ph/9402316}}.

\bibitem{Strominger:2001pn}
A.~Strominger, ``The ds/cft correspondence,'' {\em JHEP} {\bf 10} (2001) 034,
\href{http://www.arXiv.org/abs/hep-th/0106113}{{\tt hep-th/0106113}}.

\bibitem{Bird:1993yi}
{\bf HIRES} Collaboration, D.~J. Bird {\em et al.}, ``Evidence for correlated
  changes in the spectrum and composition of cosmic rays at extremely
  high-energies,'' {\em Phys. Rev. Lett.} {\bf 71} (1993)
3401--3404.

\end{thebibliography}\endgroup
\bibliographystyle{utphys}

\end{document}